\documentstyle[sprocl]{article}

\def\Journal#1#2#3#4{{#1} {\bf #2}, #3 (#4)}

\def\CQG{\em Class. Quantum Grav.}

\def\PRL{\em Phys. Rev. Lett.}
\def\PRD{{\em Phys. Rev.} D}

\def\be{\begin{equation}}
\def\ee{\end{equation}}
\def\bea{\begin{eqnarray}}
\def\eea{\end{eqnarray}}

\begin{document}

\title{GEODESIC COMPLETENESS OF DIAGONAL $G_2$
METRICS}

\author{ L. FERN\'ANDEZ-JAMBRINA }

\address{Departamento de  Ense\~nanzas B\'asicas de la Ingenier\'{\i}a Naval, ETSI
Navales, Universidad Polit\'ecnica de Madrid, Arco de la Victoria s/n,\\
E-28040-Madrid, Spain}

\maketitle\abstracts{
In this talk a sufficient condition for a diagonal orthogonally
transitive cylindrical $G_2$ metric to be geodesically complete is given. The
condition is weak enough to comprise all known diagonal perfect fluid
cosmological models that are non-singular.}
  
\section{Introduction}

The interest on non-singular cosmological models  has been triggered by the
appearance of Senovilla's cylindrical solution \cite{Seno} of Einstein's
equations. However, whereas it is easy to check the regularity of the curvature
invariants, it is usually cumbersome to determine whether a spacetime is
geodesically complete
\cite{Chinea}.

Therefore it would be appealing to have a sufficient condition on the metric
coefficients that could easy to check in order to settle the issue. 

In this talk we provide a condition that is not too restrictive in the sense
that all known non-singular diagonal cylindrical perfect fluid models are
comprised in it.

\section{Geodesic equations}

From the beginning we shall restrict ourselves to diagonal cylindrical
orthogonally transitive models. The metric can be written as,
\begin{eqnarray}
ds^2=e^{2\,g(t,r)}\left\{-dt^2+dr^2\right\}+\rho^2(t,r)e^{2\,f(t,r)}d\phi^2
+e^{-2\,f(t,r)}dz^2,\label{metric}
\end{eqnarray}
in a coordinate patch where the time  and radial coordinates  are 
isotropic and the angular and axial coordinates are
adapted to the Killing fields. The
usual ranges for the cylindrical coordinates are assumed,
\begin{eqnarray}
-\infty<t,z<\infty,\quad 0<r<\infty,\quad 0<\phi<2\,\pi.
\end{eqnarray}

The metric functions will be taken to be $C^2$ in order to have a well
defined Riemann curvature and we shall also assume that there is an
axis in the spacetime. The coordinates are chosen so that it is located  on
$r=0$.

 Since there is a two-dimensional group of isometries, the order of two of the
geodesic equations,
\begin{equation} 
\ddot x^i+\Gamma^i_{jk}\dot x^j\dot x^k = 0,\end{equation}
can be lowered by the introduction of constants of motion, $L$ and $P$, 

\begin{eqnarray}
L=\rho^{-2}(t,r)e^{-2\,f(t,r)}\dot\phi,
\end{eqnarray}

\begin{eqnarray}
P=e^{2\,f(t,r)}\dot z,
\end{eqnarray}
and the other equations can be written in a compact way as,
\begin{eqnarray}
\{e^{2\,g(t,r)}\dot t\}^{\cdot} -\frac{e^{-2g(t,r)}}{2}\left\{e^{2g(t,r)}\left[\delta
+P^2e^{2f(t,r)}+L^2\frac{e^{-2f(t,r)}}{
\rho^{2}(t,r)}\right]\right\}_t=0,
\label{eeq1}
\end{eqnarray}

\begin{eqnarray}
\{e^{2\,g(t,r)}\dot
r\}^{\cdot}+\frac{e^{-2g(t,r)}}{2}\left\{e^{2g(t,r)}\left[\delta
+P^2e^{2f(t,r)}+L^2\frac{e^{-2f(t,r)}}{
\rho^{2}(t,r)}\right]\right\}_r=0,\label{eeq2}
\end{eqnarray}
where the dot stands for derivation with respect to the affine parameter.

Finally, the equation that determines the affine parametrization is,
\begin{eqnarray}
\delta=e^{2\,g(t,r)}\left\{\dot
t^2-\dot
r^2\right\}-L^2\rho^{-2}(t,r)e^{-2\,f(t,r)}-P^2e^{2\,f(t,r)},\label{delta}
\end{eqnarray}
where $\delta$ is zero for null, one for timelike and
minus one for spacelike geodesics, that is, it is non-negative for causal
geodesics.

The previous system of equations 
can be reduced to a first order one in an efficient way by making use of
hyperbolic functions of a function
$\xi$, that is related to the radial speed along the geodesic,

\begin{eqnarray}
\dot
t(t,r)=e^{-2g(t,r)}F(t,r)\,\cosh\xi(t,r),
\end{eqnarray}

\begin{eqnarray}
\dot r(t,r)=e^{-2g(t,r)}F(t,r)\,\sinh\xi(t,r),
\end{eqnarray}

\begin{eqnarray}
\dot\xi(t,r)=-e^{-2g(t,r)}\left\{F_t(t,r)\sinh\xi(t,r)+F_r(t,r)\cosh\xi(t,r)\right\},
\end{eqnarray}

\begin{equation}
F(t,r)=e^{g(t,r)}\sqrt{\delta+L^2\frac{e^{-2f(t,r)}}{\rho^{2}(t,r)}+P^2e^{2f(t,r)}},
\end{equation}
for future-pointing geodesics. 

In order to have a similar expression for
past-pointing geodesics one has only to reverse the sign of the time derivatives
and of $\dot t$.

If we impose on this system of equations conditions in order to prevent
arbitrarily large growth of the time and radial coordinate for finite affine
parameter we get the following theorem,

\begin{description}
\item[Theorem:] A cylindrically symmetric diagonal metric in the form of Eq.~\ref{metric} with
$C^2$ metric coefficients  $f,g,\rho$ is future causally geodesically complete if the following
conditions are fulfilled for large values of $t$.
\end{description}

\begin{itemize}
\item For $\dot r>0$,
\begin{eqnarray}\label{Mxi1}
\left.
\begin{array}{l}g_r+g_t\\
(g-f-\ln\rho)_r+(g-f-\ln\rho)_t\\
(g+f)_r+(g+f)_t\end{array}\right\}>0,\end{eqnarray}

\begin{eqnarray}\label{Mxi2}
\left.
\begin{array}{l}{g_r}\\(g-f-\ln\rho)_r+(g-f-\ln\rho)_t\\(g+f)_r+(g+f)_t\end{array}\right\}
\end{eqnarray}
positive or at most of the same order as the respective terms in the previous equations. 

\item For $\dot r<0,$ 

\begin{eqnarray}
&\delta\{g_t-g_r\}+L^2\frac{e^{-2f}}{\rho^2}\left\{(g-f-\ln\rho)_t-(g-f-\ln\rho)_r\right\}+\nonumber\\&
P^2e^{2f}\left\{
(g+f)_t-(g+f)_r\right\}>0,\label{mxi1}
\end{eqnarray}

\begin{eqnarray}\delta
g_r+L^2\frac{e^{-2f}}{\rho^2}\left(g_r-f_r-\frac{\rho_r}{\rho}\right)+P^2e^{2f}\left(g_r+f_r\right),\label{mxi2}
\end{eqnarray}
negative or at most of the same order as the term in the previous equation.

\item There must be constants $a$, $b$, such that,
\begin{eqnarray}
\left.\begin{array}{c}2g(t,r)\\g(t,r)+f(t,r)+\ln\rho\\g(t,r)-f(t,r)\end{array}\right\}\ge-
\ln|t+a|+b.\label{tt}
\end{eqnarray}
\end{itemize}

A similar theorem is obtained for past-pointing geodesics just reversing the sign of the time
derivatives and expressing the conditions for small values of $t$ instead of for large values.

If the conditions in this theorem are fulfilled, the spacetime is globally
hyperbolic \cite{Geroch}, since every null geodesic intersects once and only once every
hypersurface $t=const.$. 

\section{Discussion}

The only known non-singular cylindrical diagonal perfect fluid cosmological models that are
known to us are those in \cite{Seno}, \cite{Ruiz}, \cite{Sep}, \cite{leo}. It is easy to check
that all of them fulfill the conditions stated in the theorem.  Therefore it cannot be
considered too restrictive. 

A similar condition is being prepared for non-diagonal models and it is expected to be
published soon \cite{leo2}.   

\section*{Acknowledgments}
The present work has been supported by Direcci\'on General de
Ense\~nanza Superior Project PB95-0371. The authors wish to thank
 F.J. Chinea, L.M. Gonz\'alez-Romero and M.J. Pareja Garc\'{\i}a for valuable
discussions.

\end{document}